\documentclass[usenatbib]{mn2e}
\usepackage{graphicx}
\usepackage{url}
\usepackage{caption}
\usepackage{footnote}
\usepackage{booktabs}
\usepackage{threeparttable}
\usepackage{epsfig}
\usepackage{cite}
\usepackage{hyperref}
\begin{document}

\title[Tidal tails of interacting galaxies]
{NGC 6845: metallicity gradients and star formation in a complex compact group
\thanks {Observing run: GS-2011B-Q-36
{\dag} : dolave@udec.cl}
}
\author[D. Olave-Rojas et al.]
{
\parbox[t]{\textwidth}{
 D. Olave-Rojas$^{1}$\dag, S. Torres-Flores$^{1}$, E. R. Carrasco$^{2}$, C. Mendes de Oliveira$^{3}$, D. F. de Mello$^{4}$ \& S. Scarano Jr$^{5}$,
}
\vspace*{6pt} \\
$^{1}$Departamento de F\'isica y Astronom\'ia, Universidad de La Serena, Av. Cisternas 1200, La Serena, Chile\\
$^{2}$Gemini Observatory/AURA, Southern Operations Center, Casilla 603, La Serena, Chile\\
$^{3}$ Instituto de Astronomia, Geof\'isica e Ci\^encias Atmosf\'ericas da Universidade de S\~{a}o Paulo,\\
 Cidade Universit\'aria, CEP:05508-900, S\~{a}o Paulo, SP, Brazil\\
$^{4}$ Catholic University of America, Washington, DC 20064, USA\\
$^{5}$ Departamento de F\'isica - CCET, Universidade Federal de Sergipe, Rod. Marechal Rondon s/n, 49.100-000,\\ 
Jardim Rosa Elze, S\~{a}o Cristov\~{a}o, SE, Brazil\\}
\date{\today}

\pagerange{\pageref{firstpage}--\pageref{lastpage}}

\maketitle

\label{firstpage}

\begin{abstract}
We have obtained Gemini/GMOS spectra of 28 regions located across the interacting group NGC 6845, spanning from the inner regions of the four major galaxies (NGC 6845A, B, C, D) to the tidal tails of NGC 6845A. All regions in the tails are star-forming objects with ages younger than 10 Myr. We derived the gas-phase metallicity gradients across NGC 6845A and its two tails and we find that these are shallower than those for isolated galaxies. NGC 6845A has a gas-phase oxygen central metallicity of \mbox{12+log(O/H)$\sim$8.5} and a flat gas-phase metallicity gradient ($\beta$=0.002$\pm$0.004 dex kpc$^{-1}$) out to $\sim$4 $\times$ R$_{25}$ (to the end of the longest tidal tail). Considering the mass-metallicity relation, the central region of NGC 6845A displays a lower oxygen abundance than the expected for its mass. Taking into account this fact and considering the flat oxygen distribution measured along the eastern tidal tail, we suggest that an interaction event has produced a dilution in the central metallicity of this galaxy and the observed flattening in its metal distribution. We found that the star formation process along the eastern tidal structure has not been efficient enough to increase the oxygen abundances in this place, suggesting that this structure was formed from enriched material.

\end{abstract}


\section{Introduction}\label{intro}
Local interacting/merging systems provide us with ideal laboratories to study the effect of tidal forces in the kinematic, morphology and chemical evolution of galaxies. (\citealt{toomre}, \citealt{schweizer78}). In particular, several gas-rich interacting galaxies present lower nuclear metallicities and shallower metallicity gradients than non-interacting galaxies of similar masses, suggesting that large scale gas flows may be linked to the chemical evolution of these systems (\citealt{kewley10}, \citealt{rupkea},b, \citealt{werk}, \citealt{bresolin12}).

A recent systematic survey of gas-phase metallicity gradients of a large sample of nearby non-interacting disk galaxies \citep{moran} showed that metallicity gradients within the optical radii of galaxies are flat for massive galaxies while metallicities decline steadily with radius for galaxies with low stellar mass (\mbox{log(M$_{*}$) $<$ 10.2}). Other studies have derived metallicity gradients of interacting, warped, minor mergers or paired galaxies (\citealt{kewley10}, \citealt{werk}, \citealt{rich}) and found shallower profiles than generally found in galaxies of similar mass. Recently, \citet{rosa} using Gemini data found that oxygen gradients are flatter for pairs of galaxies than for isolated spiral galaxies. These works were focused on the study of metallicity gradients of interacting galaxies up to twice the optical radii of the galaxies but none of them dealt with galaxies with tidal tails. In an attempt to probe metallicities at increasingly larger radii, our group has focused on deriving metallicity gradients for galaxies with long tidal tails in the optical and H{\sc i} as is the case of the galaxies NGC 92 \citep{torres14} and NGC 2782 (\citealt{werk}, \citealt{torres12}). The presence of tidal tails allows probing metallicities not only to much outer radii than in cases of galaxies with no tails, but also in the intragroup medium, where intergalactic H{\sc ii} regions are found (as it is the case for Stephan's quintet, \citealt{trancho}). Our previous studies analyzed NGC 92 and NGC 2782, systems which display tidal tails with flat metallicity distributions. Metallicities were obtained by using the nebular spectra of several young star forming regions located along the tidal tails. These regions are metal rich, which suggests that they were born from material expelled from the galaxies involved in the interaction. 

The study of systems in extreme phases, such as when collisions happen and form tidal tails, may help elucidate processes which are rare in the nearby Universe but may have been common at high-z. Formation of new systems due to strong interactions, and subsequent evolution of several generations of these systems may enhance the metallicity of the outskirts of galaxies and intragroup medium and may be an important mechanism for driving metals from the centers of the large galaxies outward and to the intragroup medium.  One extreme case of recently formed objects due to interactions is extranuclear H$\alpha$-emitting complexes such as those found in the ultraluminous infrared galaxies (ULIRGS) sample studied by \citet{miralles}. These may be precursors of the so called tidal dwarf galaxies. They may form independent new systems or fall back onto the parent galaxies.

In order to study these environmental effects on galaxy formation and evolution we have an observational program to obtain deep imaging and spectroscopy of a sample of galaxies with tidal tails. Two groups have already been studied, NGC 92 and NGC 2782. In the current paper, we analyze the group NGC 6845,  formed by two spirals and two lenticular galaxies, NGC 6845A, B, C and D  (also known as Klemola 30 by \citealt{klemola}, and ESO284-IG008), where one of the galaxies (NGC 6845A) has long tidal tails. Here we present deep Gemini imaging and spectroscopy of the system revealing star-forming regions and yielding a map of the metal distribution across a projected distance to the center of the galaxy of 140 kpc. This group has a mean optical recession velocity of 6701 km s$^{-1}$, which has been corrected by the Virgo, Great Attractor and Shapley infall \citep{mould}. Assuming H$_0$=73 km s$^{-1}$ Mpc$^{-1}$ this velocity implies a distance to NGC 6845 of 91.8 Mpc. We note that 1 arcsec $\simeq$ 0.445 kpc at this distance.

The paper is organized as follows. In Section 2 we describe the system and the data, imaging and spectroscopy, taken with the Gemini South telescope. In Section 3 we present the data analysis. In Section 4 we present the results. In Section 5 we discuss our results and in Section 6 we present our main conclusions.

\section{System and data description}
\subsection{The NGC~6845 group}
\begin{figure}
        \includegraphics[width=\columnwidth]{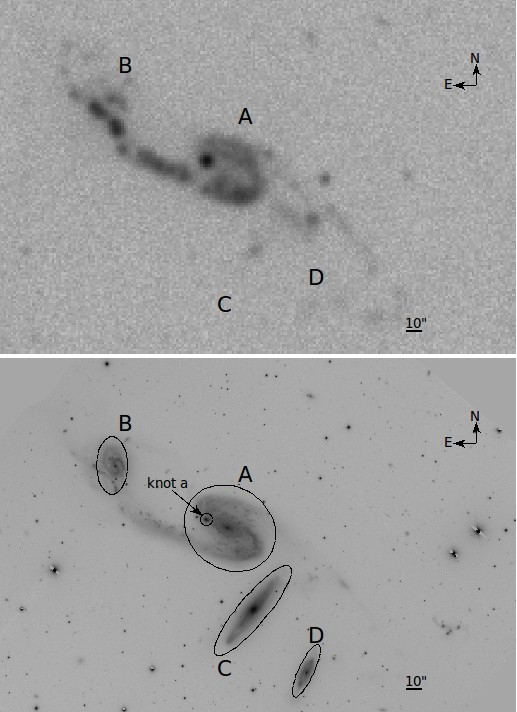}
	\caption{The top panel shows a \textit{GALEX} FUV image, obtained from the All-Sky Imaging Survey (AIS). The bottom panel shows the Gemini image in the \textit{r'}-band. The black ellipses show the optical radius (R$_{25}$) for each galaxy.}

	\label{fuv_r}
\end{figure}

The NGC 6845 group (see Fig. \ref{fuv_r}) was first studied by \citet{graham}, who reported signatures of strong interaction events between the members. These authors found that NGC 6845A has a long tidal arm projected onto NGC 6845B. They were able to confirm that NGC 6845A and NGC 6845B are part  of the system. Also, they found one exceptionally large knot in NGC~6845A - they named it as ``knot a". Later, NGC 6845C and NGC 6845D were confirmed as members of the group by \citet{rose}. Deep images by \citet{rose} showed the bright eastern arm of NGC 6845A extending towards NGC 6845B and a tidal bridge bluer than the inner disk of NGC 6845A.

\citet{rodrigues} found that the tidal bridge between NGC~6845A and NGC~6845B has a \textit{(B-I)} colour bluer than that of the inner disk of NGC~6845A. These authors found eight strong condensations, identified as H{\sc ii} regions with ages between 3-8Myrs. They also studied the blue knot ``a", but they named it ``7". 

\citet{gordon} obtained H{\sc i} (21-cm) observations of NGC~6845 by using the Australia Telescope Compact Array (ATCA), during 1997 and 1998. The beam size was 43 $\times$ 36 arcsec$^{2}$, with a channel width of 20 km s$^{-1}$. Using these data, \citet{gordon} derived integrated H{\sc i} intensity maps, mean velocity field and H{\sc i} velocity dispersion maps. In addition, 20-cm radio continuum maps were derived from these observations. \citet{gordon} found that the H{\sc i} emission is associated with the galaxies NGC 6845A and NGC 6845B and that it has a typical column density of N$_{HI}$$\sim$18.3$\times$10$^{20}$cm$^{-2}$. The far-ultraviolet (FUV) Galaxy Evolution Explorer (\textit{GALEX}) image of this group shows that the galaxies NGC~6845A and NGC~6845B have strong UV emission and galaxies NGC~6845C and NGC~6845D have very weak or null UV emission (see top panel in Fig. \ref{fuv_r}).

The optical and morphological properties of the galaxies in the NGC 6845 quartet are summarized in Table \ref{group_op}. Total magnitudes in the B-band and total \textit{(B-V)} colors for all members of the group were obtained from the HyperLeda data base \citep{maka}. Sizes, H{\sc i} masses and inclinations for the main members of NGC 6845 were taken from literature (see Table \ref{group_op}).

\begin{table*}
\begin{threeparttable}
	\caption{Optical and morphological properties for the galaxies in the NGC 6845 quartet}
  	\label{group_op}
        {\small
 		\begin{tabular}{lcccc}
		\hline
		\hline
 		Galaxy     	  &  NGC 6845A                    & NGC 6845B	                    & NGC 6845C                     & NGC 6845D	              \\
		\hline
Center position\tnote{\emph{a}} & 20:00:58.1, -47:04:12           & 20:01:05.9, -47:03:35           & 20:00:56.6, -47:05:02           & 20:00:53.4, -47:05:38           \\
Size [arcmin$^2$]\tnote{\emph{a}} & 3.7 $\times$ 1.7               & 1.1 $\times$ 0.6               & 0.6 $\times$ 0.2               & 0.8 $\times$ 0.5               \\
              \textit{i} [degree] & 57$\pm$3\tnote{\emph{b}}             &  0$\pm$5\tnote{\emph{b}}             &  77$\pm$3\tnote{\emph{b}}            & 71$\pm$2\tnote{\emph{b}}             \\
         M$_{HI}$\tnote{\emph{a}} & (1.8$\pm$0.2)$\times$10$^{10}$M$_\odot$  & (5$\pm$2)$\times$10$^{8}$M$_\odot$   &      -                         & -                              \\  
    M$_{B}$ [mag]\tnote{\emph{c}} & 13.83 $\pm$ 0.60               & 14.90 $\pm$ 0.28               & 16.35 $\pm$ 0.33               & 15.43 $\pm$ 0.28               \\  
      (B-V) [mag]\tnote{\emph{c}} & 0.61                           & 1.01                           & 0.92                           & 0.74                           \\

		\hline
		\hline
		\end{tabular}  
\begin{tablenotes}
\item[\emph{a}]{The center position (J2000.0), size and H{\sc i} mass were obtained from \citet{gordon}}
\item[\emph{b}]{The inclination was obtained from \citet{rodrigues}}
\item[\emph{c}]{Total magnitudes in the B-band and \textit{(B-V)} colors were obtained from HyperLeda data base \citep{maka}}
\end{tablenotes}
}
\end{threeparttable}
\end{table*}

\subsection{Data}
Images and spectra for NGC 6845 were taken with the Gemini MultiObject Spectrograph (GMOS \citealt{hook}) at the Gemini South telescope under the science program GS-2011B-Q-36 (PI: S. Torres-Flores).

We used these data to determine the oxygen abundances of the objects located in the tidal tails of NGC 6845A and between NGC 6845A and NGC 6845B.

\subsubsection{Imaging}\label{imagedata}
Gemini GMOS images were taken in \textit{u'}, \textit{g'} and \textit{r'}-band filters with exposure times of 5$\times$ 300s, 4$\times$ 300s, and 3$\times$ 300s, respectively. All images were acquired during the night of 2011-09-24 UT Date with an instrumental position angle of 50$^{\rm o}$, a 2$\times$2 binning mode, a mean seeing of 0.62'', 0.68'' and 0.40'', and airmass 1.2, 1.1 and 1.1 for the \textit{u'}, \textit{g'} and \textit{r'}-band filters, respectively. The Gemini images have a field of view of 5.5$\times$5.5 arcmin$^{2}$ and a pixel scale (binned) of 0.146 arsec/pixel. 

Individual images were corrected by bias and flat-field and the astrometry was performed using the {\sc theli} software \citep{schirmer}. We produced a master bias using five bias frames. We also generated several master flats (one for each filter) using five flat-fields for each filter. These bias and flat-field images were then used to correct each science frame. Finally the science frames were then combined using the option median with sigma clipping.

The astrometry and co-addition for each filter and the multiple frames were performed with the {\sc theli} software. The sky in all images was normalized using a constant sky value. The zero-point photometric calibrations were derived from one field only, the  photometric standard field \mbox{040020-600200}, which was observed in the  \textit{u'}, \textit{g'} and \textit{r'}-band filters. The field was observed under photometric conditions in the same night of the science images and in similar observing conditions. Calibration data were reduced in the same way as science imaging data. In order to obtain instrumental magnitudes, we performed aperture photometry by using the task {\sc phot}, in the {\sc noao} package in {\sc iraf}, where these magnitudes were corrected by aperture. Then, instrumental magnitudes were converted into calibrated magnitudes by applying a linear transformation that includes the air mass and extinction coefficient for each filter and the standard magnitudes, which were obtained from the \textit{Southern Standard Stars for the u'g'r'i'z' System catalogue} \citep{smith02}. We have fixed the extinction coefficients using the coefficients listed in the GMOS web pages. This procedure allow us to obtain the zero-point for each filter, which are 24.34$\pm$0.10, 27.79$\pm$0.02 and 27.74$\pm$0.01 for the filters \textit{u'}, \textit{g'} and \textit{r'} respectively.

\subsubsection{Spectroscopy}
\begin{figure*}
        \includegraphics[width=1.\textwidth]{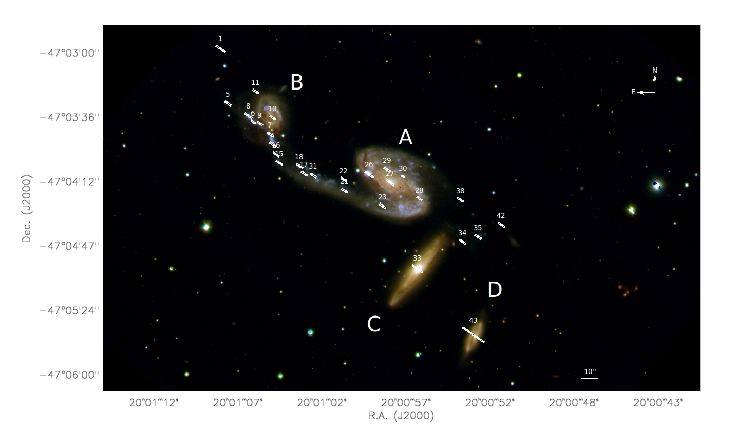}
	\caption{Gemini/GMOS colour composite image (blue-\textit{u'} filter, green-\textit{g'} filter and red-\textit{r'} filter) of NGC 6845.  White rectangles indicate the regions with spectroscopic data.}
	\label{spec1}
\end{figure*}

Given that we are interested in determining the oxygen abundance gradient along the tidal tail of NGC 6845A, several requirements had to be taken into account when selecting the spectroscopic targets. For instance, we needed to observe star-forming regions located at different positions of the tail, in order to increase the radial coverage. In addition, the sources had to contain emission lines, given that we estimate oxygen abundances by using strong-line methods. This last point was usually accomplished when we selected blue and bright sources. Finally, the GMOS mask design included another criterion to select our targets, given that each slit had to have a minimum length in order to allow for sky subtraction. For all these reasons, the spectroscopic targets were selected by visual inspection, in order to increase the number of useful sources to determine the abundance gradient. In Fig. \ref{spec1} we show a false colour Gemini/GMOS image of NGC 6845 where the white boxes indicate the regions for which we have obtained spectroscopic data. In order to study the properties of the main galaxies of the system, we have placed slits over the centers of these objects.

\begin{figure*}
\centering
        \includegraphics[width=0.7\textwidth]{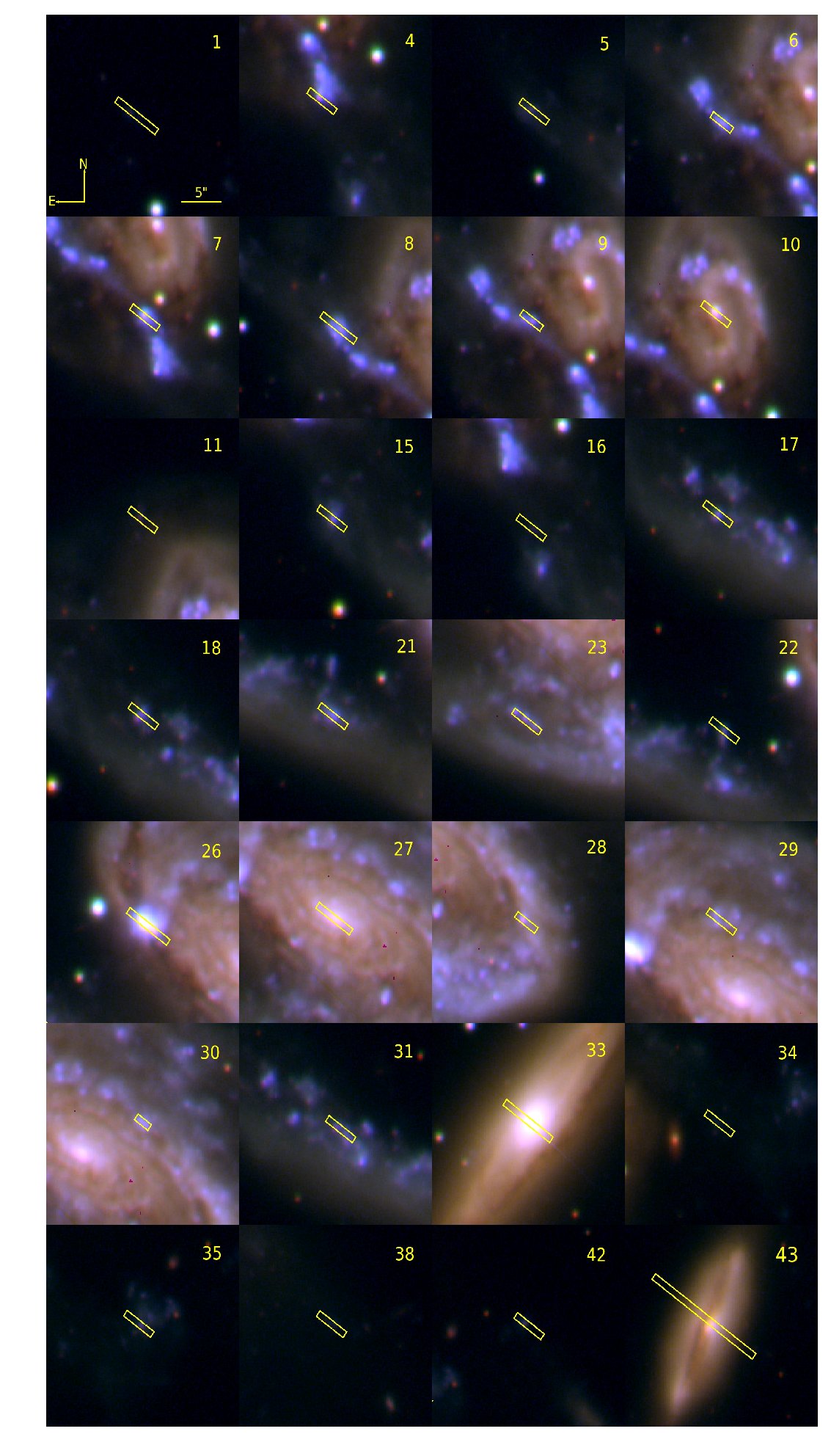}
	\caption{Gemini/GMOS colour composite image (blue-\textit{u'} filter, green-\textit{g'} filter and red-\textit{r'} filter) of the star-forming regions and the central regions of the main galaxies (rectangles) in the observed area of NGC 6854.}
	\label{clumps}
\end{figure*}

We have designed one GMOS mask using the \textit{r'} filter image to obtain the spectroscopy of the selected regions. Since our sources are mostly point-like we adopted slit widths of 0.75'' to maximize the amount of incoming light, given that this matched the expected seeing. The slit lengths were set individually with the purpose of improving the number of observed star-forming regions (see Fig \ref{clumps}). In general, slit lengths ranged between 1.8 and 10.5 arc sec (except for the main galaxies which needed longer slit lengths). All slits were oriented according to the instrumental position angle of 50$^{\rm o}$. We used the R400\_G5325 grism and obtained four exposures of 1200s in three different central wavelengths (one in 5750{\AA}, two in 5800{\AA} and one in 5850{\AA}). This setup resulted in a spectral coverage of 4015--8018 {\AA}. These values were chosen in order to detect H$\beta$, [O{\sc iii}]$\lambda$5007 {\AA}, H$\alpha$, and [N{\sc ii}]$\lambda$6584 {\AA} lines, which are used to estimate the oxygen abundances. The mean seeing during the spectroscopic observations was 0.5'' and the spectral resolution was 4.1 {\AA} at H$\alpha$ (R$\sim$1500). All spectroscopic data were obtained in 2012-09-16 UT Date.

The flat-field frames were observed at the same position of the target, with mean airmass of 1.06, 1.08, 1.11 and 1.14, for the central wavelengths of 5800, 5750, 5850 and 5800 {\AA}, respectively. The CuAr arc lamps were observed as day time calibrations. Given that the CuAr arc lamps were not observed in the same sequence as the science observations, flexures can introduce some systematic uncertainties in the wavelength calibration. When an offset in the spectral direction due to flexure was detected, this was then corrected using {\sc gstransform} task. These uncertainties do not affect the fluxes but will affect the errors in the radial velocity measurements. Due to the chosen angle (see Section \ref{imagedata}) and by not following the recommendation to use the parallax angle some of the light from the object could be lost out of the slit, but because of the low air mass of the observations, this effect is minimized and considered negligible.

The GMOS spectroscopic observations were reduced using the {\sc Gemini} package from {\sc iraf}. The four frames were corrected for overscan, bias and flat-field using the routines {\sc gbias}, {\sc gsflat} and {\sc gsreduce}. The cosmic rays were removed using {\sc lacos\_spec} \citep{lacos}. Individual 2D spectra of each slit, for each central wavelength observed, were combined to remove the detector gaps and the remaining  cosmic rays using the {\sc gemcombine} task. The wavelength calibration was performed with the task {\sc gswavelenght} and {\sc gstransform} and the rms in the wavelength calibration was 0.16 {\AA}. The sky lines were removed from the spectra using the task {\sc gsskysub}. Following similar procedures, we have reduced the spectrum of the standard star LTT9239, which was observed to determine the sensitivity function in our observations and to flux calibrate all spectra by using the task {\sc gsstandard} and {\sc gscalibrate}, respectively. This star was observed under the science program GS-2012B-Q-60 in 2012 September. The above process allowed us to obtain a calibrated unidimensional spectrum for each region in this study, which is analyzed as described below. We note that the instrumental setup used to observe this standard star was the same defined for the observations of NGC 6845.

\section{Analysis}
\subsection{Photometry}
We have done aperture photometry in the \textit{u'}, \textit{g'} and \textit{r'}-band filters for the 24 star-forming regions where we have centered the slits. The different magnitudes were measured by using the task {\sc phot}, in {\sc iraf}, where we used the zero-point magnitudes listed in section \ref{imagedata}. For each filter we used an aperture diameter of 0.75'', which matches the slit width we chose for the spectroscopy (we note that the remaining four regions for which we have spectroscopy correspond to the center of the main galaxies). The aperture of 0.75'' in diameter was chosen to compare the results derived from the photometry and spectroscopy using the same physical area for the selected region. Using a common aperture, we can perform a comparison between the results derived from photometry and spectroscopy thus obtaining more robust and accurate conclusions. The sky was obtained with the option mode, over an annulus with an inner radius of 2.5" (17 pixels, which is 3.3 times the aperture diameter) and a width of 1" or (7 pixels), corresponding to the parameters {\sc annulus} and {\sc danullus} in {\sc phot}, respectively. For regions \#28, \#29 and \#30, located on the disk of NGC 6845A, the sky values were estimated from zones located outside the galaxy in a region without stellar emission. At the end, we produce a catalogue with the \textit{u'}, \textit{g'} and \textit{r'}-band magnitudes (and their associated uncertainties) for each star-forming region (see Table \ref{flux}).  Magnitudes were corrected by Galactic extinction using A$_u$=E(B-V)$\times$4.21, $A_g$=E(B-V)$\times$3.27, $A_r$=E(B-V)$\times$2.27, respectively, where the extinction coefficients were taken from \citet{fitz} and the E(B-V) values were taken from \citet{schlegel}.

\begin{figure}
\centering
        \includegraphics[width=\columnwidth]{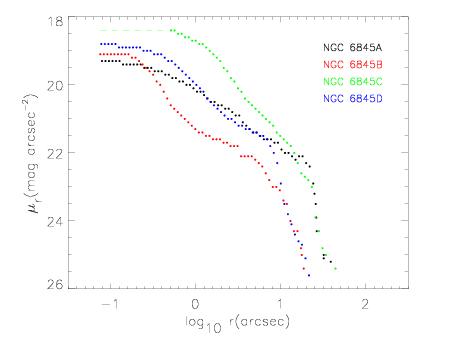}
	\caption{Surface brightness profiles for each galaxy derived by using the task {\sc ellipse}}
	\label{perfil}
\end{figure}

We have derived surface brightness profiles for the main galaxies of the system NGC 6845. These profiles were obtained by using the task {\sc ellipse} in {\sc iraf} and we use the \textit{r'}-band image. This task fits elliptical isophotes to galaxy images and the input parameter are the initial isophote center, semi-major axis length, ellipticity and position angle. The output is a catalogue with values for the semi-major axis length and mean isophotal magnitude. Mainly, this procedure was developed in order to obtain the optical radius (R$_{25}$) of each galaxy (see Fig. \ref{fuv_r}). In Fig. \ref{perfil} we show the surface brightness profiles for the different members of NGC 6845.

\subsection{Spectroscopy}\label{anspec}

All observed spectra were corrected for Galactic extinction using the {\sc idl} code {\sc fm\_unred} \citep{fitz}. We use the absolute extinction value of A(V)=0.124 from S\citet{sha}.”

\subsubsection{Main galaxies}\label{maingal}
We used the {\sc emsao} task in the {\sc rvsao} package inside {\sc iraf} to measure the radial velocities of all members. This task uses absorption/emission line features in an observed spectrum to determine the radial velocity. In the case of spectrum dominated by emission lines, each line is fitted with a simple Gaussian. The central value of the Gaussian is compared with the rest frame wavelength of the emission line and then the radial velocity is computed. The Gaussian fitting is performed for all emission lines presented in the observed spectrum simultaneously. The same procedure can be used for spectrum dominated by absorption line features. This is the case of NGC 6845C, whose radial velocity was estimated using the most prominent absorption line features. We have summarized the heliocentric radial velocities (v$_{hel}$) for the galaxies in the NGC 6845 quartet in Table \ref{group_vel}. The {\sc emsao} task measures the uncertainties in the radial velocities through a cross-correlation between emssion lines in the input spectrum and the catalogue. 

In Table \ref{group_vel} we have also included the heliocentric velocities measured by \citet{rose} for the different members of NGC 6845. Comparing our measurements with the values reported by \citet{rose} we detected a difference of the order of $\Delta_{vel}$$\sim$ 50 km s$^{-1}$ for NGC 6845A and $\Delta_{vel}$$\sim$ 30 km s$^{-1}$ for NGC 6845B, C and D. Considering the uncertainties, there is a good agreement for the velocities of members B, C and D. The main difference in velocity comes from member NGC 6845A. Possible reasons for this difference can be related with the number of lines used to estimate the velocity (6 lines versus 8 lines in our case). In addition, \citet{rose} used longslit observations, therefore, the aperture could have been defined in a slightly different way than it was done for our multislit. Also the wavelength calibration, depending on the location of the slit, may affect measurements causing slightly different values.

The spectra of the main galaxies of NGC 6845 are dominated by absorption and emission lines. Some emission lines can be affected by absorption features, which are associated with an underlying stellar population, e.g. in the case of the H$\beta$ line. Therefore, emission line fluxes were measured once the stellar continuum was subtracted. This procedure was performed by using the Penalized Pixel-Fitting (p{\sc pxf}) code \citep{capelari}, which allowed us to model the stellar continuum, which was subtracted from the observed spectra. This analysis allowed us to obtain a pure emission line spectra for each galaxy of the group NGC 6845. We note that we applied this procedure just to remove the continuum emission.

\begin{table}
\centering
\begin{threeparttable}
	\caption{Velocities of the galaxies in the NGC 6845 quartet}
  	\label{group_vel}
	{\small
 		\begin{tabular}{lcc}
		\hline
		\hline
 		Galaxy      & v$_{hel}$\tnote{\emph{a}} & v$_{hel}$\tnote{\emph{b}}  \\
                            &\multicolumn{2}{c}{km s$^{-1}$}                                                      \\
		\hline
		NGC 6845A  &  6463$\pm$22  & 6410$\pm$0   \\
		NGC 6845B  &  6805$\pm$30  & 6776$\pm$4   \\
		NGC 6845C  &  6785$\pm$12  & 6755$\pm$17  \\
		NGC 6845D  &  7036$\pm$47  & 7070$\pm$43  \\
		\hline
		\hline
		\end{tabular}  
\begin{tablenotes}
\item[\emph{a}]{This paper}
\item[\emph{b}]{\citet{rose}}
\end{tablenotes} 
}
\end{threeparttable}
\end{table}

\subsubsection{Star forming-regions}\label{specsf}

The emission-line fluxes for the observed spectra were measured with the {\sc splot} task in the {\sc noao} package in {\sc iraf}. The uncertainties in the fluxes were estimated through a Monte Carlo simulation inside of the {\sc splot} task, where a Gaussian model is fitted to the emission lines. The fluxes and their uncertainties are summarized in Table \ref{flux}.

We have carefully inspected all the spectra in order to search for absorption features around the H$\beta$ emission line. We found that regions \#4, \#26 and \#29 display a weak absorption around H$\beta$. As suggested by the referee, we have fitted a Gaussian on this absorption, in order to recover all the emission at this location. We found that the measured value (absorption in H$\beta$) in each case is quite low. In fact, these measurements are of the same order than the uncertainty in emission fluxes. Clearly, this confirmation make the absorptions negligible. 

In some cases we detect H$\gamma$ emission line (about 10 regions). We have used that emission line (in addition to H$\alpha$ and H$\beta$) to correct our data for internal extinction. We found that the use of the H$\gamma$ give us a higher internal extinction. In order to correct these 10 spectra, we used the average of the internal extinction derived from H$\gamma$ with the extinction derived from H$\alpha$ and H$\beta$. Finally, we estimated the fluxes and oxygen abundances for the regions and we compare these new estimations with the estimates obtained originally (corrected by using H$\alpha$ and H$\beta$). We found that the oxygen abundances were similar. The larger differences  in the oxygen abundances were of the order of 0.01 dex (and just in two regions). For this reason (and for a homogeneous analysis), we kept our correction just by using the H$\alpha$ and H$\beta$ emissions lines, given that this was the method that we applied for all regions and given that the scientific results are exactly the same.

In the case of the emission line dominated spectra, each spectrum was corrected for internal reddening through the Balmer decrement using the H$\alpha$ over H$\beta$ line ratio. The value for the intrinsic H$\alpha$/H$\beta$ ratio was taken from \citet{oster} for T$_e$=10000 K and N$_e$ =100 and the correction by internal reddening was done using the {\sc idl} code {\sc calz\_unred} \citep{calzeti}. This extinction law works for the wavelength interval ranging from 1200 to 22000 {\AA} and it has a ratio of total to selective extinction of R$_{V}$=4.05$\pm$0.80, which was obtained by using optical-infrared (IR) observations of starburst galaxies  and this extinction law has a typical uncertainty of 2 per cent \citep{calzeti}. In practice, we provide the observed spectra and a colour excess, and the {\sc calz\_unred} code apply the correction. In this case the colour excess corresponds to the stellar colour excess, which correlates with the gaseous reddening by \mbox{E(B-V)$_{star}$=0.44 $\times$E(B-V)$_{gas}$.}  In just two spectra we have not detected the H$\beta$ line and in these cases we have used the same procedure as that in \citet{torres14}. We adopted the median value of the extinction in the nearest regions for which it was possible to calculate the H$\alpha$/H$\beta$ ratio. The internal gaseous colour excess measured through the Balmer decrement for the star-forming regions located in the  NGC 6845 group have values ranging from E(B-V)=0.00 to 0.77. These values are in agreement with the results found by  other authors for star-forming regions in interacting galaxies (see \citealt{smith}).

The coordinates, corrected emission-line fluxes, radial velocities and colour excess measured for all the star-forming regions belonging to NGC 6845 are presented in Table \ref{flux}. Radial velocities were estimated with the {\sc emsao} task, in {\sc iraf}.

The spectra of most of the star-forming regions show a flat continuum with a combination of strong nebular emission and null absorption lines. One exception is region \#26, which clearly shows a blue stellar continuum. 

\begin{table*}
\centering
\begin{minipage}[t]{\textwidth}
\scriptsize
\caption{Position, magnitude, flux, velocities, distance and colour excess of the region in NGC 6845}
\begin{tabular}{cccccccccccc}
\hline
        ID  & R.A.  &  Dec.  & r'\footnote{Magnitudes were measured in a fixed aperture diameter of 0.75 arcsec and corrected by MW extinction and no internal extinction.} & H$\beta$  & [O{\sc iii}]$\lambda$5007  & H$\alpha$ & [N{\sc ii}]$\lambda$6584  & Velocity    & Distance\footnote{De-projected distance to the center of galaxy NGC 6845A}  & E(B-V)\footnote{Internal gaseous extinction for each source. These values were estimated following \citet{dominguez}} & E(B-V)\footnote{Stellar extinction for each source. These values were estimated following \citet{calzeti}} \\
                        &\multicolumn{2}{c}{(J2000.)} & (mag) & \multicolumn{4}{c}{($\times$ 10$^{-15}$erg cm$^{-2}$ s$^{-1}$ \AA$^{-1}$)} & (km s$^{-1})$ & (kpc) & \multicolumn{2}{c}{(mag)}	  \\
          	\hline		
		   1    &  20:01:08.186 & -47:02:53.75 & 24.04$\pm$0.02 &  0.25$\pm$0.11 & 0.35$\pm$0.12 & 0.69$\pm$0.10  & 0.11$\pm$0.05 & 6765$\pm$18  & 71.037 & 0.22$\pm$0.10 & 0.10$\pm$0.04 \\
		   4    &  20:01:05.299 & -47:03:49.85 & 21.18$\pm$0.01 &  0.33$\pm$0.27 & 0.92$\pm$0.25 & 1.02$\pm$0.25  & 0.14$\pm$0.09 & 6300$\pm$22  & 49.539 & 0.11$\pm$0.15 & 0.05$\pm$0.06 \\
		   5    &  20:01:07.795 & -47:03:25.88 & 23.53$\pm$0.03 &  0.07$\pm$0.03 & 0.36$\pm$0.29 & 0.32$\pm$0.11  & 0.07$\pm$0.06 & 6260$\pm$29  & 66.435 & 0.14$\pm$0.14 & 0.06$\pm$0.06 \\
		   6    &  20:01:06.300 & -47:03:36.60 & 21.75$\pm$0.02 &  0.84$\pm$0.19 & 4.73$\pm$2.47 & 3.77$\pm$0.29  & 0.40$\pm$0.16 & 6264$\pm$16  & 56.226 & 0.60$\pm$0.09 & 0.26$\pm$0.04 \\
		   7    &  20:01:05.441 & -47:03:43.77 & 20.73$\pm$0.01 &  6.32$\pm$0.30 &46.51$\pm$6.98 &30.11$\pm$4.52  & 2.27$\pm$0.24 & 6355$\pm$15  & 50.364 & 0.77$\pm$0.07 & 0.34$\pm$0.03 \\
		   8    &  20:01:06.569 & -47:03:32.75 & 21.92$\pm$0.01 &  -  	         & 0.69$\pm$0.23 & 0.62$\pm$0.21  & 0.14$\pm$0.10 & 6301$\pm$20  & 58.055 &  -	    &  -	    \\
 		   9    &  20:01:05.904 & -47:03:37.74 & 21.77$\pm$0.01 &  0.75$\pm$0.22 & 1.34$\pm$0.27 & 2.24$\pm$0.32  & 0.36$\pm$0.17 & 6324$\pm$18  & 53.489 & 0.04$\pm$0.17 & 0.02$\pm$0.07 \\
		   11   &  20:01:06.209 & -47:03:18.82 & 23.68$\pm$0.03 &  0.15$\pm$0.09 & 0.21$\pm$0.13 & 0.50$\pm$0.15  & 0.09$\pm$0.07 & 6725$\pm$18  & 56.425 & 0.02$\pm$0.17 & 0.01$\pm$0.07 \\
		   15   &  20:01:04.884 & -47:04:00.56 & 22.24$\pm$0.01 &  0.82$\pm$0.17 & 1.83$\pm$0.15 & 2.56$\pm$0.34  & 0.33$\pm$0.13 & 6355$\pm$16  & 47.263 & 0.19$\pm$0.10 & 0.08$\pm$0.05 \\
		   16   &  20:01:04.990 & -47:03:55.38 & 24.77$\pm$0.10 &  0.13$\pm$0.04 & 0.20$\pm$0.07 & 0.46$\pm$0.19  & 0.10$\pm$0.04 & 6329$\pm$28  & 47.633 & 0.33$\pm$0.11 & 0.15$\pm$0.05 \\
		   17   &  20:01:03.418 & -47:04:06.90 & 22.77$\pm$0.03 &  0.38$\pm$0.17 & 0.68$\pm$0.18 & 1.11$\pm$0.24  & 0.22$\pm$0.11 & 6321$\pm$19  & 37.067 & 0.05$\pm$0.16 & 0.02$\pm$0.07 \\
		   18   &  20:01:03.715 & -47:04:02.87 & 22.39$\pm$0.01 &  0.78$\pm$0.19 & 1.35$\pm$0.23 & 2.46$\pm$0.35  & 0.41$\pm$0.16 & 6471$\pm$16  & 38.918 & 0.10$\pm$0.15 & 0.05$\pm$0.07 \\
		   21   &  20:01:01.044 & -47:04:17.01 & 22.31$\pm$0.02 &  0.55$\pm$0.16 & 0.82$\pm$0.18 & 1.72$\pm$0.27  & 0.36$\pm$0.12 & 6365$\pm$17  & 20.604 & 0.18$\pm$0.13 & 0.08$\pm$0.06 \\
		   22   &  20:01:01.123 & -47:04:10.96 & 22.90$\pm$0.01 &  0.08$\pm$0.06 & 0.18$\pm$0.07 & 0.27$\pm$0.13  & 0.06$\pm$0.03 & 6327$\pm$36  & 20.759 & 0.16$\pm$0.16 & 0.07$\pm$0.07 \\
		   23   &  20:00:58.721 & -47:04:27.14 & 23.02$\pm$0.06 &  0.21$\pm$0.13 & 0.14$\pm$0.08 & 0.62$\pm$0.19  & 0.21$\pm$0.12 & 6462$\pm$16  &  4.391 & 0.04$\pm$0.16 & 0.02$\pm$0.07 \\
		   26   &  20:00:59.741 & -47:04:07.73 & 18.79$\pm$0.01 &  8.94$\pm$0.80 & 6.72$\pm$0.72 &33.00$\pm$2.04  &10.64$\pm$0.62 & 6134$\pm$17  & 12.911 & 0.56$\pm$0.09 & 0.24$\pm$0.04 \\
		   28   &  20:00:56.659 & -47:04:22.51 & 21.25$\pm$0.01 &  4.08$\pm$0.26 & 5.03$\pm$0.28 &16.60$\pm$1.40  & 4.80$\pm$0.33 & 6703$\pm$23  & 13.380 & 0.70$\pm$0.07 & 0.31$\pm$0.03 \\
		   29   &  20:00:58.591 & -47:04:04.71 & 21.28$\pm$0.01 &  0.13$\pm$0.06 & -  	   & 0.29$\pm$0.17  & 0.08$\pm$0.05 & 6330$\pm$26  & 12.415 & 0.24$\pm$0.21 & 0.10$\pm$0.09 \\
		   30   &  20:00:57.694 & -47:04:08.74 & 21.15$\pm$0.01 &  0.52$\pm$0.14 & 0.38$\pm$0.06 & 1.61$\pm$0.20  & 0.56$\pm$0.14 & 6423$\pm$17  & 13.212 & 0.68$\pm$0.06 & 0.30$\pm$0.03 \\
		   31   &  20:01:02.851 & -47:04:08.03 & 22.99$\pm$0.07 &  -  	         & 0.07$\pm$0.04 & 0.27$\pm$0.13  & 0.09$\pm$0.05 & 6305$\pm$16  & 33.009 &  -	    &  -	    \\
		   34   &  20:00:54.276 & -47:04:46.93 & 24.96$\pm$0.046 &  0.09$\pm$0.04 & 0.08$\pm$0.04 & 0.13$\pm$0.10  & 0.03$\pm$0.02 & 6596$\pm$46  & 26.421 & 0.04$\pm$0.15 & 0.02$\pm$0.07 \\
		   35   &  20:00:53.405 & -47:04:44.60 & 23.41$\pm$0.016 &  0.24$\pm$0.14 & 0.36$\pm$0.14 & 0.62$\pm$0.17  & 0.11$\pm$0.07 & 6635$\pm$23  & 33.162 & 0.00$\pm$0.16 & 0.00$\pm$0.07 \\
		   38   &  20:00:54.365 & -47:04:22.77 & 24.73$\pm$0.038 &  0.07$\pm$0.03 & 0.13$\pm$0.07 & 0.15$\pm$0.08  & 0.03$\pm$0.03 & 6681$\pm$33  & 30.703 & 0.07$\pm$0.12 & 0.03$\pm$0.05 \\
		   42   &  20:00:52.090 & -47:04:37.32 & 23.65$\pm$0.013 &  0.12$\pm$0.08 & 0.28$\pm$0.16 & 0.35$\pm$0.12  & 0.04$\pm$0.03 & 6696$\pm$20  & 44.540 & 0.02$\pm$0.18 & 0.01$\pm$0.08 \\
\hline
\hline
\vspace{-0.8cm}
\label{flux}
\end{tabular}
\end{minipage}
\end{table*}

\section{Results}
\subsection{Ages and internal extinctions of the star-forming regions}
Ages and internal extinctions for all regions were estimated by comparing the observed colours with respect to the colours derived from the spectrophotometric models obtained from Starburst99 (SB99, \citealt{leit}) for the \textit{u'}, \textit{g'} and \textit{r'}-band filters. We used a single stellar population model (instantaneous burst), Salpeter initial mass function (IMF) with a mass boundary of 0.1-100 M$_{\odot}$. Given that we are able to estimate the oxygen abundances of the regions, the SB99 models were generated for a metallicity of Z=0.008 (0.4$\times$Z$_{\odot}$). This value corresponds to the average of the metallicities of the different star-forming regions. Finally, ages were obtained by minimizing the $\chi^2$ between the observed colours and the models (see  \citeyear{torres14}). 

We would like to note that the star-forming regions located in the tails of NGC 6845A may be formed by stellar associations that we are not able to resolve. In some cases, these star-forming objects could be not massive enough to sample the complete IMF. This effect should affect the observed colours of these sources, producing uncertainties in the age determination of low-mass systems (given that the SB99 models considers a fully sampled IMF). This kind of effect has been reported before by other authors (e.g. \citealt{dasilva}, \citeyear{dasilva14}, \citeyear{dasilva15}).

\begin{figure}
        \includegraphics[width=\columnwidth]{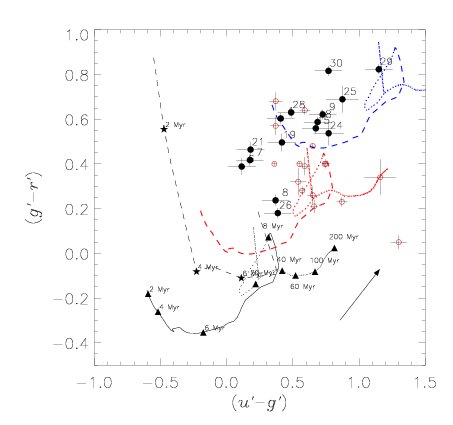}
	\caption{Gemini (\textit{u'}-\textit{g'}) versus (\textit{g'}-\textit{r'}) diagram for the star-forming regions located in the tidal tail of NGC 6845A. Solid black, red and blue dashed lines represent the SB99 models with no extinction, a starburst extinction law from \citet{calzeti94} with E(B-V)=0.3 and a starburst extinction law with E(B-V)=0.7, respectively. Red circles represent the regions detected in the tidal tail of NGC 92.}
	\label{color}
\end{figure}

The colour-colour diagram for the regions located in the NGC 6845 system is shown in Fig \ref{color} (black filled circles). Solid black, red and blue dashed lines represent the SB99 models with no extinction, a starburst extinction law from \citealt{calzeti} with E(B-V)=0.3 and a starburst extinction law with E(B-V)=0.7, respectively. We note that the H$\alpha$ emission line can have a strong contribution in the broad-band magnitudes of young star-forming objects. For instance, \citet{smith} noted that the \textit{r'}-band magnitude could be affected by 1.1 mag in the case of regions having ages of $\sim$ 1Myr. This effect is shown by a black dashed line in Fig. \ref{color}, which has been taken from \citet{smith}. For comparison, in Fig. \ref{color} we have included the regions located in the tidal tail of NGC 92 (red circles, \citealt{torres14}). Colours (\textit{u'}-\textit{g'}) and (\textit{g'}-\textit{r'}) are listed in Table \ref{magcol} together with internal extinctions and ages derived for these regions. The uncertainties in the colours were estimated considering the uncertainties in the magnitudes for each filter and the uncertainties in the zero-point magnitudes. We found that ages for most of the star-forming regions are in the range $\sim$2 to $\sim$7 Myr consistent with the value obtained by \citet{rodrigues}. The ages that we have estimated are in agreement with the spectra, which show strong emission lines, thus we can conclude that all 24 selected objects are young star-forming regions. The ages for the star-forming regions in NGC 6845 are in agreement with the results found by other authors, for similar regions. For example, \citet{smith} studied the star-forming regions in the pair of galaxies NGC 2856/4. These authors found ages in the range of $\sim$4-20 Myrs. The young ages for our star-forming regions suggest that these regions were formed in situ (\citealt{chien}, \citealt{demelloa}, \citealt{smith}).

In order to determine the colour excess of each star-forming region in NGC 6845, we have used the starburst extinction law, defined by \citet{calzeti}, using A$_u$=E(B-V)$\times$6.08, $A_g$=E(B-V)$\times$4.737, $A_r$=E(B-V)$\times$3.54, respectively. The internal extinctions (i. e. the color excesses) measured for the star-forming regions in the NGC 6845 system have values E(B-V)=0.50-0.94 that agree with the values found in star forming regions located in other interacting systems e.g. NGC 2856, NGC 2782 and NGC 92 (see \citealt{smith}, \citealt{torres12}, \citeyear{torres14}). 

\begin{table}
\centering
\begin{threeparttable}
	\caption{Colour and colour excess of the star-forming regions located in the tidal tail of NGC 6845A}
  	\label{magcol}
        {\small
 		\begin{tabular}{cccrr}
		\hline
		\hline
 		Region    & (\textit{u'}-\textit{g'})  & (\textit{g'}-\textit{r'})     & E(B-V)\tnote{\emph{a}} & Ages\tnote{\emph{b}}	 \\
		    ID    & \multicolumn{3}{c}{mag}     &  Myr	   \\ 
		\hline
		       4.  &  0.41 $\pm$ 0.10	&	 0.60 $\pm$ 0.02	   &  0.70	 &   2. $\pm$  1. \\
		       6.  &  0.11 $\pm$ 0.10	&	 0.39 $\pm$ 0.04	   &  0.50	 &   2. $\pm$  0. \\
		       7.  &  0.18 $\pm$ 0.10	&	 0.42 $\pm$ 0.02	   &  0.54	 &   2. $\pm$  1. \\
		       8.  &  0.37 $\pm$ 0.10	&	 0.24 $\pm$ 0.02	   &  0.50	 &   5. $\pm$  3. \\
		       9.  &  0.73 $\pm$ 0.10	&	 0.62 $\pm$ 0.02	   &  0.82	 &   4. $\pm$  3. \\
		      15.  &  0.67 $\pm$ 0.10	&	 0.56 $\pm$ 0.02	   &  0.76	 &   4. $\pm$  3. \\
		      18.  &  0.69 $\pm$ 0.10	&	 0.59 $\pm$ 0.02	   &  0.78	 &   4. $\pm$  3. \\
		      19.  &  0.42 $\pm$ 0.11	&	 0.50 $\pm$ 0.04	   &  0.66	 &   3. $\pm$  2. \\
		      21.  &  0.18 $\pm$ 0.11	&	 0.46 $\pm$ 0.03	   &  0.56	 &   2. $\pm$  0. \\
                      24.  &  0.77 $\pm$ 0.11	&	 0.54 $\pm$ 0.05	   &  0.76	 &   5. $\pm$  3. \\
		      25.  &  0.87 $\pm$ 0.12	&	 0.69 $\pm$ 0.06	   &  0.88	 &   5. $\pm$  3. \\
		      26.  &  0.39 $\pm$ 0.10	&	 0.18 $\pm$ 0.02	   &  0.46	 &   5. $\pm$  4. \\
		      28.  &  0.49 $\pm$ 0.10	&	 0.63 $\pm$ 0.02	   &  0.74	 &   3. $\pm$  2. \\
                      29.  &  1.15 $\pm$ 0.10	&	 0.82 $\pm$ 0.02	   &  0.62	 &   7. $\pm$  7. \\
		      30.  &  0.77 $\pm$ 0.10	&	 0.82 $\pm$ 0.02	   &  0.94	 &   3. $\pm$  3. \\

		\hline
		\hline
		\end{tabular}  
\begin{tablenotes}
\item[\emph{a}]{Internal E(B-V) estimated from the photometry}
\item[\emph{b}]{Ages estimated from (\textit{u'}-\textit{g'}) and (\textit{g'}-\textit{r'}).}
\end{tablenotes}
}
\end{threeparttable}
\end{table}

\subsection{Excitation mechanisms of the star-forming regions}

\begin{figure*}
        \includegraphics[width=0.8\textwidth]{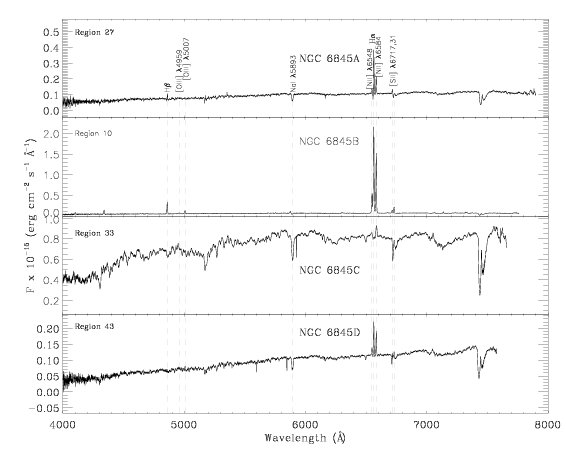}
	\caption{Gemini spectra of NGC 6845A, B, C and D (from top to bottom panel) The main emission and absorption lines are labelled.}
	\label{spec_group}
\end{figure*}

As shown in Fig. \ref{spec_group}, the main galaxies of the NGC 6845 group have typical spectra which agree with what one would expect from their morphological features. The late-type spiral galaxies have spectra with emissions lines and few absorption lines and the early-type galaxies show spectra with several absorption lines and few emission lines.

\begin{figure}
        \includegraphics[width=\columnwidth]{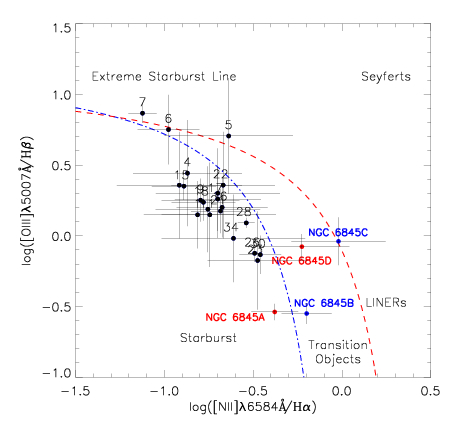}
	\caption{BPT diagram \citep{bald} for the regions located in NGC 6845. The blue and red dashed line corresponds to the limit between H{\sc ii} regions and AGNs, as described in \citet{kauf} and \citet{kew}, respectively. We also included the central regions for NGC 6845A, B, C and D.}
	\label{bpt}
\end{figure}

With the purpose to confirm and analyze the excitation mechanisms of the main galaxies and star-forming regions in NGC 6845 we have used the Baldwin–Phillips–Terlevich (BPT) diagnostic diagram \citep{bald}, which is shown in Fig. \ref{bpt}. In Fig. \ref{bpt} we have plotted the [O{\sc iii}]/H$\beta$ versus the [N{\sc ii}]/H$\alpha$ line ratios and in this figure blue and red dashed line corresponds to the limit between star-forming objects and active galactic nuclei
 (AGN), as described in \citet{kauf} and \citet{kew}, respectively.

Taking into account the uncertainties in the flux, we found that most of the 24 young star-forming regions are placed in the starburst area of the diagram, except for regions \#5, \#6 and \#7. These regions, located in the extreme starburst area of the diagram, are within a complex region of the system where the tidal arm of NGC 6845A reaches the outskirts of NGC 6845B and is projected onto it.

According to Fig. \ref{bpt}, we can see that NGC 6845A is located in the starburst area. NGC 6845B and D are located in the transition zone between the starburst and the low-ionization nuclear emission-line region (LINER), however, the uncertainties are large in both cases. Finally, NGC 6845C is located in the limit between star-forming objects and AGN proposed by \citet{kew}. As in the previous cases, the uncertainties can place this object as in the transition zone as in the AGN region.

Some authors as \citet{stasinska} have studied the composite zone between AGN and star-forming objects. They found that this category includes real AGN hosts and a fraction of non-AGN galaxies whose old stellar populations mimic the line ratios produced by AGN radiation. Also, \citet{vitale} found that the nature of the composite zone is still unclear because this region displays a mixture of AGN and star-forming galaxies. To clarify the nature of this region, \citet{scott} studied a sample of interacting galaxies. These authors found that AGN activity may increase due to galaxy interactions. Moreover, \citet{scoville} studied the star-forming regions in M51 and the nuclear starburst in Arp 220 to determine if a link between nuclear starburst and AGN can be established. This author suggests that the gas in the central regions is essential to trigger a nuclear starburst activity or light up the black hole in the galaxy. In addition, \citet{elison} studied a sample of 11060 galaxies with a companion from Sloan Digital Sky Survey (SDSS). These authors found an enhancement in the AGN fraction in interacting galaxies, which was consistent with previous results. In this sense, we could expect AGN activity in NGC 6845.

\begin{figure}
        \includegraphics[width=\columnwidth]{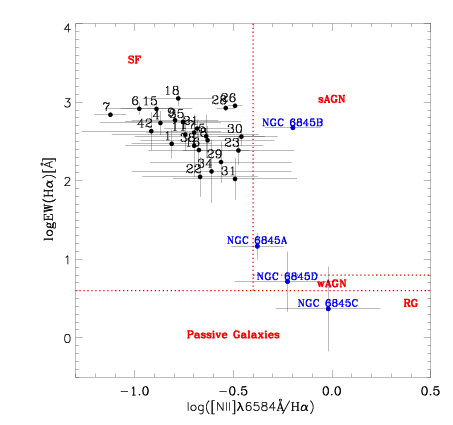}
	\caption{WHAN diagram \citep{cid11} for regions located in NGC 6845. We also included the central regions for NGC 6845A, B, C and D.}
	\label{whan}
\end{figure}

Given that the transition zone can be populated with real AGN and non-AGN host, we have used a second diagnostic diagram to elucidate the origin of the ionization mechanism in the main galaxies of NGC 6845. This diagram was proposed by \citet{cid11}, and it allows to classify galaxies by using the [NII]/H$\alpha$ ratio versus the equivalent width of the H$\alpha$ emission line (WHAN diagram). In this diagram it is possible to separate the different ionization mechanism, i.e., young stars, AGN activity and ionization produced by hot low-mass evolved stars. This last phenomenon can be associated with the \textit{retired galaxies}, i.e. galaxies on which there is no more star formation, but there is a presence of emission lines (\citealt{stasinska}, \citealt{cid11}). In Fig. \ref{whan} we show the [NII]/H$\alpha$ versus EW(H$\alpha$) for all objects for which we have spectroscopic data, where SF, sAGN, wAGN and RG define star-forming objects, Seyfert AGN, weak AGN and retired galaxies, respectively. The EW(H$\alpha$) values were measured with the task {\sc splot} in {\sc iraf}. We found that all star-forming regions observed in the whole group are located in the Star Formation locus, as expected from the BPT diagram shown in Fig. \ref{bpt}. The case of the main galaxies of the system is slightly different. According to the uncertainties in the [NII]/H$\alpha$ ratio, the galaxy NGC 6845A is located between the SF and the sAGN region. On the other hand, NGC 6845B is a sAGN. Considering this fact and its position on the BPT diagram, we can not exclude that in this object there is a contribution from a real AGN. In the case of NGC 6845C, the WHAN diagram suggests that this object could be a retired galaxy, where the ionization comes from evolved low-mass stars (however, the error bars do not allow us to be conclusive). Finally, NGC 6845D is located in the wAGN region. This fact is consistent with its position ion the BPT diagram shown in Fig. \ref{bpt}. However, the WHAN diagram could suggest that the ionization in this object is not produced by hot low-mass evolved stars, suggesting that this object could be a LINER. Considering our results, we can speculate that the nuclear region of NGC 6845A, B and D can be ionized partially by AGN activity.

\subsection{Radial velocities and radial distances}\label{ra_vel}
\begin{figure*}
\centering
        \includegraphics[width=0.9\textwidth]{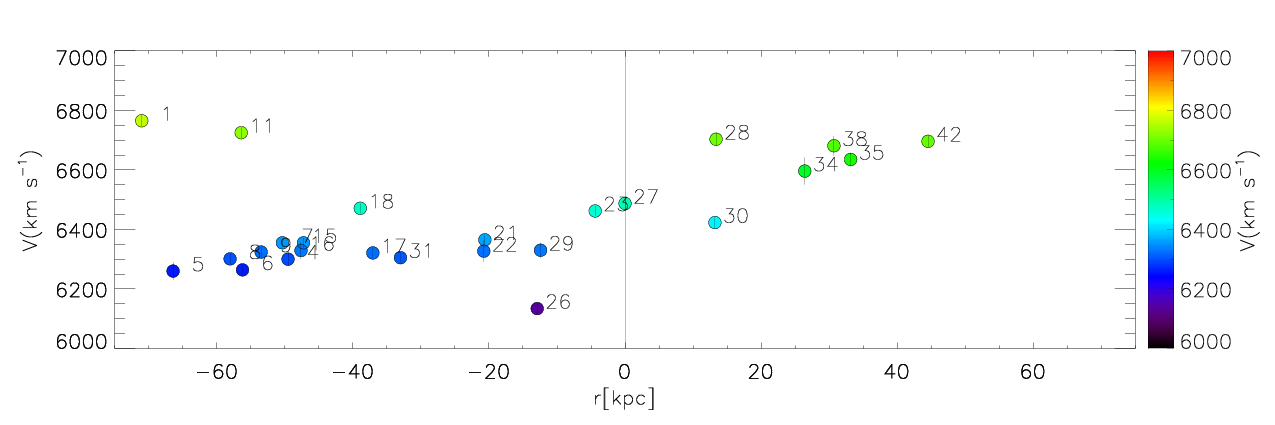}
	\caption{Radial velocity distribution along both tidal tails of NGC 6845A. Note that regions \#1 and \#11 have velocities similar to that of NGC 6845B and are unlikely to be part of NGC 6845A}
	\label{vel}
\end{figure*}

In order to determine the metallicity gradient for NGC 6845A, we need to know which regions are linked with the tidal tail of NGC 6845A, and therefore, we have measured the radial velocities for each star-forming region. To do this we used the same procedure we used for the main members of the system (see \S \ref{maingal}). The radial velocity distribution along the tidal tail of NGC 6845A (Fig. \ref{vel}) reveals that regions \#1 and \#11 have velocities typical of NGC 6845B (v$_{rad}$$\sim$6800 km s$^{-1}$) and probably are not part of NGC 6845A. For this reason, these two regions were not included in the metallicity fits provided below.

Once we know which regions are linked to NGC~6845A, it is necessary to estimate the de-projected distance from each star-forming region to the center of this galaxy. For this purpose we used the procedure described by \citet{scarano}. These authors used the position angle, inclination and the coordinates of the center of the galaxy and position of each star-forming region to deproject distances to the center of the main galaxy. We have adopted the following values: the position angle is 63$^o$ taken from \citet{gordon} and the inclination is 57$^o$ taken from \citet{rodrigues}. Our results are listed in Table \ref{flux} (see column Distance).

Finally, the optical radius of the main galaxies were derived by using the task {\sc ellipse}, in {\sc iraf}, as described above. We found that the optical radius of NGC 6845A is R$_{25}\sim$ 0.5 arcmin or $\sim$14 kpc in the \textit{r'}-band.

\subsection{Oxygen abundances and metallicity gradients}

We have estimated the oxygen abundances using semi-empirical methods. In this way, our estimations can not be understood as absolute values due to the different values which each individual calibrator provides. In this sense each individual calibrator can be used to determine relative abundances for a given object. This analysis has been used by other authors to determine the abundance of oxygen and in the study of metallicity gradients \citep[e.g.][]{bresolin12}.

Oxygen abundances for all regions were estimated using the calibrators: N2 and O3N2. These calibrators are defined as 

\begin{equation}\label{n2}
N2 = \log  \left(\frac{[N{\sc II}]\lambda6584\mathring{A}}{H\alpha}\right)
\end{equation}

\begin{equation}\label{o3n2}
O3N2 = \log  \left(\frac{([O{\sc III}]\lambda5007\mathring{A}/H\beta)}{([N{\sc II}]\lambda6584\mathring{A}/H\alpha)}\right)
\end{equation}

The N2 and O3N2 were recently re-calibrated by \citet{marino}.These authors used spectral data from the Calar Alto Legacy Integral Field Area (CALIFA) Survey and from the literature \citep{sanchez} to determine the direct oxygen abundance and also to compare with semi-empirical methods. These authors found a dispersion of 0.16 and 0.18 dex in their calibration for the N2 and O3N2, respectively (see equations 2 and 4 from \citealt{marino}).

In Table \ref{oxy} we have summarized the results for the oxygen abundance of the star-forming regions observed in NGC 6845. The uncertainties in the oxygen abundances were estimated including the uncertainties in the fluxes and the dispersion derived from N2 and O3N2 calibrations (see \citealt{marino}). Inspecting these values we found that most of the regions in NGC 6845 have sub-solar oxygen abundances (for comparison, we have added the solar abundance value given by \citealt{allende}. They found a value of 12+log(O/H)$_{\odot}$=8.69). 

Following the N2 index proposed by \citet{marino} our results are in the range between 12+log(O/H)=8.32$\pm$0.23 and 12+log(O/H)=8.52$\pm$0.16. Our values are lower than the values found by \citet{chien} for the northern tidal tail of the merging galaxy pair NGC 4676. These authors found values for the oxygen metallicity in the range 12+log(O/H)= 8.86-8.97, but they used the R$_{23}$ method. In addition, this difference in oxygen abundances may also be related with the different masses of the galaxies studied by \citet{chien} and the masses of the systems studied in this paper.

\begin{table}
\begin{threeparttable}
	\caption{Oxygen abundances for the star-forming regions located in the tidal tails and NGC 6845 system}
  	\label{oxy}
	{\small
 		\begin{tabular}{lccccc}
		\hline
		\hline
 		ID          & N2 & O3N2 & 12+log(O/H)\tnote{\emph{a}} & 12+log(O/H)\tnote{\emph{b}} \\
		            &    &      &                             &                             \\
		\hline
 1  & -0.81$\pm$0.20  & 0.96$\pm$0.31 &   8.37$\pm$0.19  &  8.33$\pm$0.19    \\
 4  & -0.87$\pm$0.31  & 1.31$\pm$0.50 &   8.34$\pm$0.22  &  8.25$\pm$0.21    \\ 
 5  & -0.64$\pm$0.36  & 1.34$\pm$0.55 &   8.45$\pm$0.23  &  8.25$\pm$0.22    \\
 6  & -0.98$\pm$0.17  & 1.73$\pm$0.30 &   8.29$\pm$0.18  &  8.16$\pm$0.19    \\
 7  & -1.12$\pm$0.08  & 1.99$\pm$0.11 &   8.22$\pm$0.17  &  8.11$\pm$0.18    \\
 8  & -0.63$\pm$0.34  &  -            &   8.45$\pm$0.23  &  -		     \\
 9  & -0.80$\pm$0.22  & 1.05$\pm$0.27 &   8.38$\pm$0.19  &  8.31$\pm$0.19    \\
11  & -0.74$\pm$0.37  & 0.89$\pm$0.52 &   8.40$\pm$0.24  &  8.34$\pm$0.21    \\
15  & -0.89$\pm$0.18  & 1.24$\pm$0.21 &   8.33$\pm$0.18  &  8.27$\pm$0.19    \\
16  & -0.68$\pm$0.26  & 0.88$\pm$0.33 &   8.43$\pm$0.20  &  8.35$\pm$0.19    \\
17  & -0.70$\pm$0.24  & 0.96$\pm$0.33 &   8.42$\pm$0.20  &  8.33$\pm$0.19    \\
18  & -0.78$\pm$0.18  & 1.02$\pm$0.22 &   8.38$\pm$0.18  &  8.32$\pm$0.19    \\
21  & -0.68$\pm$0.17  & 0.86$\pm$0.23 &   8.43$\pm$0.18  &  8.35$\pm$0.19    \\
22  & -0.67$\pm$0.29  & 1.03$\pm$0.48 &   8.43$\pm$0.21  &  8.31$\pm$0.21    \\
23  & -0.48$\pm$0.28  & 0.30$\pm$0.45 &   8.52$\pm$0.21  &  8.47$\pm$0.20    \\
26  & -0.49$\pm$0.04  & 0.37$\pm$0.07 &   8.52$\pm$0.16  &  8.45$\pm$0.18    \\
28  & -0.54$\pm$0.05  & 0.63$\pm$0.06 &   8.49$\pm$0.16  &  8.40$\pm$0.18    \\
29  & -0.56$\pm$0.36  &  -            &   8.48$\pm$0.23  &  -		     \\
30  & -0.46$\pm$0.12  & 0.33$\pm$0.18 &   8.53$\pm$0.17  &  8.46$\pm$0.18    \\
31  & -0.49$\pm$0.31  &  -            &   8.52$\pm$0.22  &  -		     \\
34  & -0.61$\pm$0.40  & 0.59$\pm$0.51 &   8.46$\pm$0.25  &  8.41$\pm$0.21    \\
35  & -0.76$\pm$0.29  & 0.94$\pm$0.42 &   8.39$\pm$0.21  &  8.33$\pm$0.20    \\
38  & -0.70$\pm$0.35  & 1.00$\pm$0.47 &   8.42$\pm$0.23  &  8.32$\pm$0.21    \\
42  & -0.92$\pm$0.35  & 1.27$\pm$0.52 &   8.32$\pm$0.23  &  8.26$\pm$0.21    \\
                \hline
		\hline
		\end{tabular}  
\begin{tablenotes}
\item[\emph{a}]{Oxygen abundances estimated by using the N2 calibrator proposed by \citet{marino}}
\item[\emph{b}]{Oxygen abundances estimated by using the O3N2 calibrator proposed by \citet{marino}}
\end{tablenotes}
}
\end{threeparttable}
\end{table}

\begin{figure*}
\centering
        \includegraphics[width=0.95\textwidth]{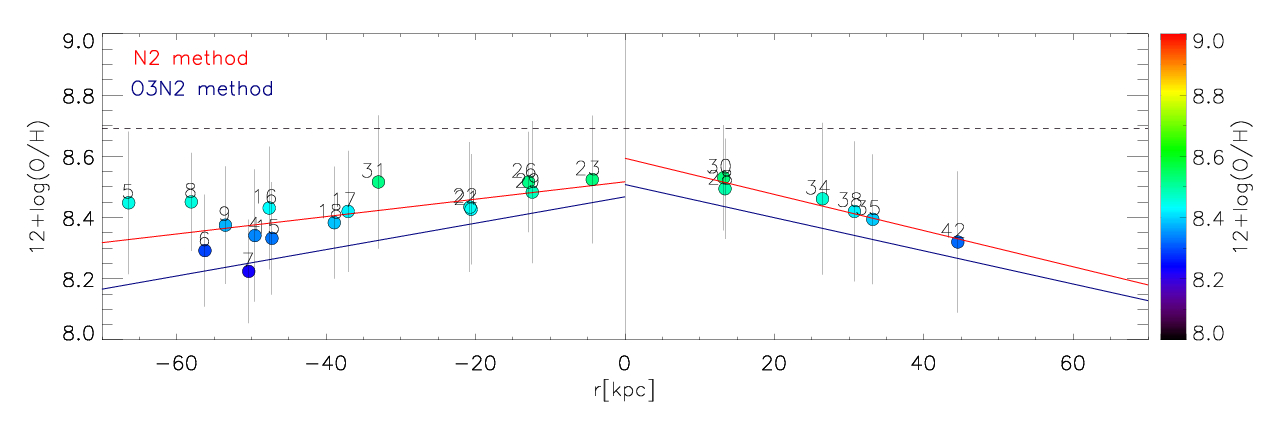}
	\caption{Galactocentric oxygen abundance gradients for NGC 6845A. Solid red and blue lines represent a linear fit for N2 and O3N2 index, respectively. The dashed horizontal line shows the solar oxygen abundance 12+log (O/H)=8.69 \citep{allende}. Note that fit values are provided in Table \ref{ajusab}).}
	\label{grad}
\end{figure*}

Because a couple of regions do not display the H$\beta$ emission line (sources \#8 and \#31), the metallicity gradient and conclusions obtained in this paper were made using the values derived through the N2 index proposed by \citet{marino}. The N2 and O3N2 calibrators provided a similar trend in the oxygen abundance distribution for the system NGC 6845A (in the sense of the slope of the fit). The only difference between the gradient derived with these calibrators was a small shift in the the zero-point (see Table \ref{ajusab}), which is shown in Figure \ref{grad}.

Fig. \ref{grad} shows the distribution of oxygen abundances of the star forming regions located across the disk of NGC 6845A and of those between NGC 6845A and NGC 6845B and between NGC 6845A and NGC 6845C (see Fig \ref{spec1}) using the N2 index for the aforementioned reason. For comparison we have added the fit obtained using the O3N2 index. The points in the plot represent the oxygen abundance for the star-forming regions obtained following the N2 index. The star-forming regions lie towards  slightly systematically low O/H values as a function of distance from the NGC 6845A nucleus. The slope and zero points of the radial oxygen abundances were obtained by using the {\sc idl} routine {\sc mpfitexy} \citep{williams}. The {\sc mpfitexy} code determines the linear fits to the data set with errors in both variables. The uncertainties in the deprojected distances were estimated considering the pixel scale in the images (see \S \ref{imagedata}) and the uncertainty of the distance to the source. In Table \ref{ajusab} we show the zero-points and slopes of the fits. The tidal tail between NGC 6845A and NGC 6845B has a slope of $\alpha_{(A-B)}$=0.003$\pm$0.001 dex kpc$^{-1}$ and the tidal tail between NGC 6845A and NGC 6845C has a gradient of $\alpha_{(A-C)}$=-0.006$\pm$0.001 dex kpc$^{-1}$ both measured using the N2 method proposed by \citet{marino}. The value for the slope in the fits between NGC 6845A and NGC 6845B is similar to that found for NGC 92 ($\alpha$=-0.0017$\pm$0.0021, see \citealt{torres14}). On the other hand, the value for the slope in the fits between NGC 6845A and NGC 6845C is shallower than the value found by \citet{rupkeb} for the mean result obtained for the control sample used in their study ($\beta_{c}$=-0.041$\pm$0.009) and is even shallower than the mean value of interacting galaxies in their sample (\mbox{$\beta_{i}$=-0.017$\pm$0.002}).

\begin{figure*}
\centering
        \includegraphics[width=0.95\textwidth]{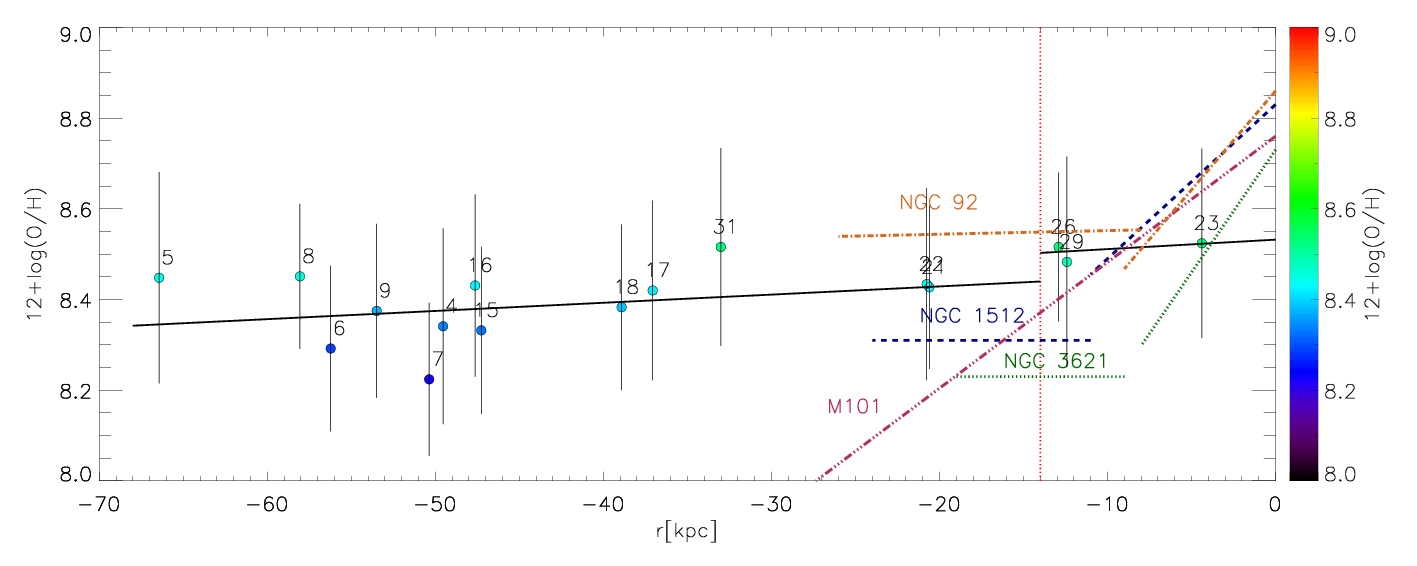}
	\caption{Galactocentric oxygen abundance gradients for NGC 6845A following the N2 index calibrated by \citet{marino} and the schematic representation for NGC 92, NGC 1512, NGC 3621 and M101 (\citealt{torres14}, \citealt{bresolin12} and \citealt{keni03}). We have made a fit for the inner (R $<$ R$_{25}$) and outer regions (R $>$ R$_{25}$) separately. The break in the solid black line is located in R$_{25}$ and is represented by a red dotted vertical line (R$\sim$14kpc) (fit values are provided in Table \ref{ajusab}). Regions in the area between NGC 6845A and C are not included in the fit.}
	\label{gradiente}
\end{figure*}

Given that one of the main goals of this paper is the study of the tidal tail that connects NGC 6845A and NGC 6845B, for convenience and simplicity, Fig. \ref{gradiente} shows the abundance gradient of the northeast side of NGC 6845A covering 66 kpc from the nucleus of NGC 6845A. The solid black line represents the fit made for the oxygen abundance following the N2 index calibrated by \citet{marino}. We have made a fit for the inner (R $<$ R$_{25}$) and outer regions (R $>$ R$_{25}$) separately. The break in the solid black line is located in R$_{25}$ (R$\sim$14kpc). For comparison, we added the schematic representation of the galactocentric oxygen abundance gradients obtained in the literature for NGC 1512, NGC 3621 \citep{bresolin12}, M101 \citep{keni03}, \citep{breso07} and NGC 92 \citep{torres14}. The oxygen abundances for these galaxies were estimated using the N2 method calibrated by \citet{petini}. As expected, the gradients of NGC 6845A are significantly different from normal disks (M101). However, the slope of the abundance gradient for NGC 6845A is very different from NGC 1512 which also has clear signs of tidal interactions. The slope of the abundance in the inner disk (R $<$ R$_{25}$) of NGC 6485A is flatter than that for NGC 1512, and roughly flat in the outer disk (R $>$ R$_{25}$). In Table \ref{ajusab} we listed the zero-points and slopes. According to Fig. \ref{gradiente} and considering the uncertainties (see table \ref{ajusab}) we can see a smooth transition in the oxygen abundance gradients between the inner (R $<$ R$_{25}$) and outer (R $>$ R$_{25}$) regions of NGC 6845A ($\beta$=0.002$\pm$0.030 and $\beta$=0.002$\pm$0.004, respectively). We note that due to the low number of internal regions (R $<$ R$_{25}$) observed, the determination of the inner slope is uncertain. Considering the uncertainties, the slope for the outer regions is similar to the value found for NGC 92 (see \citealt{torres14}). 

 \begin{table*}
\centering
	\caption{Zero-points and slopes of the linear fits for the tidal tails  in NGC 6845A}
  	\label{ajusab}
	{\small
 		\begin{tabular}{lcccc}
		\hline
              	\hline
                Tail between             &\multicolumn{2}{c}{NGC 6845A-NGC 6845B} & \multicolumn{2}{c}{NGC 6845A-NGC 6845C} \\
		Index                    & Zero Point    & Slope                     & Zero Point	& Slope 	   \\
                                         & 12+log(O/H)   & dex kpc$^{-1}$            &  12+log(O/H)	& dex kpc$^{-1}$    \\
		\hline
                \vspace{0.1cm}
                N2    & 8.52$\pm$0.11 & 0.003$\pm$0.003          & 8.59$\pm$0.19 & -0.006$\pm$0.007    \\
                \vspace{0.1cm}
                O3N2 & 8.47$\pm$0.13 & 0.004$\pm$0.003        & 8.51$\pm$0.20 & -0.005$\pm$0.007    \\
                \hline
                \vspace{0.1cm}
                NGC 6845A (R$<$R$_{25}$):&               &                         &  &                       \\
                \vspace{0.1cm}
		N2    & 8.53$\pm$0.33 & 0.002$\pm$0.030  & -& - \\
		\vspace{0.1cm}
                O3N2 & 8.48$\pm$0.32 & 0.002$\pm$0.032  & - & - \\
		\hline
                \vspace{0.1cm}
                NGC 6845A (R$>$R$_{25}$): &                &                  & &   \\
                \vspace{0.1cm}
		N2    & 8.47$\pm$0.19 & 0.002$\pm$0.004 &- &-  \\
		\vspace{0.1cm}
                O3N2 & 8.41$\pm$0.21 & 0.003$\pm$0.004  & - & - \\
		\hline
		\hline
                \vspace{-0.8cm}
		\end{tabular}}
\end{table*}

\citet{moran} found that the average metallicity profile of 174 star-forming galaxies is strikingly flat out to the radius enclosing 90\% (R$_{90}$) of the r-band light. Given that we just have two regions inside R$_{90}$ (\#23 and \#29), we can not be conclusive in comparing the gradient inside R$_{90}$.

\section{Discussion}
\subsection{Possible scenarios for the metal distribution in NGC 6845A: gas mixing hypothesis}
During the last years several simulations have shown that gas inflows in merging/interacting galaxies can trigger strong bursts of star formation \citep[e.g.][]{perez}. With this process, we should expect to observe a dilution in the nuclear metal abundance of these systems \citep{rupkea}. Several physical phenomena can affect the evolution of the nuclear metallicity in interacting/merging systems. For example, \citet{torrey} suggest that during the first pericenter passage and the final merger, the central metallicities of the interacting/merging galaxies can be affected by radial inflows of less enriched gas (due to tidal interactions). In addition, these authors also suggest that the central chemical abundances can be increased by star formation events and also they claim that the galactic winds can play a role in the chemical evolution of the nuclear regions of interacting/merging galaxies. These authors found that the  initial fraction of gas plays a fundamental role in the evolution of the nuclear metallicity. What happens in the outer regions of galaxies is not fully understood either in the extended disk of interacting or non-interacting systems. For example, \citet{bresolin12} studied two galaxies with interaction signatures. One of these objects, NGC 1512, displays an extended star formation, as seen in its UV image, this is the same scenario that NGC 6845 shows (see fig. \ref{fuv_r}). The metallicity gradient derived is negative out to R$_{25}$. After that, it becomes flat. \citet{bresolin12} suggest that accretion of gas coming from the intergalactic medium (IGM), which should be already enriched, can produce this phenomenon. Also, metal transport from the centre to the outer region could be a possible explanation. Recently \citet{lopez15} studied the system NGC 1512. These authors suggest that a merger event (which already took place in this galaxy several Gyrs ago) could increase the metallicities in the outskirts of this galaxy. However, some authors have found that star-forming regions located in tidal tails display super solar and solar oxygen abundances and flat metallicity distributions (\citealt{chien} and \citealt{torres14}, respectively). \citet{chien} argued that the flat distributions may suggest gas mixing within the tail. However, \citet{torres14} did not detect gas flows in the tidal tail of NGC 92 (through the use of kinematics from H$\alpha$ Fabry-Perot data). These findings suggest that the metal mixing process already took place several Myrs ago and now we are seen the current star formation, which is based on pre-enriched material. This scenario is supported by the work developed by \citet{avillez}, who found that the mixing time-scales are short (on the scale of Myrs).

In the case of NGC 6845A, what is the best scenario to explain its observed oxygen distribution abundance? In the inner region (R$<$R$_{25}$), we measured a slope of \mbox{$\beta$=0.002$\pm$0.030 dex kpc $^{-1}$}. This value indicates a flat metal distribution when compared with non-interacting systems (see \citealt{zaritsky}). In the case of the outer region, we obtained a value of \mbox{$\beta$=0.002$\pm$0.004 dex kpc$^{-1}$}. This value is similar (considering the uncertainties) to the slope of the metal distribution found in the tail of NGC 92 (\mbox{$\beta$=-0.0008$\pm$0.0032 dex kpc $^{-1}$}). Moreover, the tail of NGC 6845A is more extended than that of NGC 92 (as shown in Fig. \ref{gradiente}). These results are in agreement with the scenario on which galaxy interactions flatten metallicity gradients. In this context, and inspecting Table \ref{ajusab}, we can consider that NGC 6845A has a central metallicity of 12+log(O/H)$\sim$8.5-8.6. If an interaction event has produced the observed flattening in the metallicity gradient, this process eventually should have diluted the central abundance of NGC 6845A. In order to check this point, we have inspected the mass-metallicity relation published by \citet{tremonti}, where the abundances where estimated using photoionization models. \citet{lopez15} note that these models could give abundances that are $\sim$0.3-0.4 dex higher than the values obtained from the direct method, which could be comparable with abundances derived from strong line methods. Considering this point, NGC 6845A should have a stellar mass of \mbox{log (M$_{\star}$)$\sim$9.5-10.2 M$_{\odot}$}. However, that mass range is even lower than the \textit{actual HI content} of this galaxy, which corresponds to \mbox{log (M$_{\sc HI}$)=10.25 M$_{\odot}$ \citep{gordon}}. In addition, we have estimated the stellar mass of NGC 6845A, by using its H-band absolute magnitude (taken from the NED data base), their (B-V) optical colour (listed in Table 1), and considering the relations given in \citet{bell}. This exercise gives us a stellar mass of \mbox{log (M$_{\star}$)$\sim$10.86 M$_{\odot}$}. In this sense, the stellar mass of NGC 6845A is larger than the value expected from the mass metallicity relation. Another important and interesting relation of studying is the luminosity-metallicity relation.

In addition, \citet{kew06} studied the luminosity-metallicity relation (B-band absolute magnitude versus oxygen abundance) for a sample of interacting galaxies. These authors compare their results with a control sample composed by field galaxies, finding that close pairs display lower central abundances than field galaxies. We have converted the B-band magnitude of NGC 6845A (see Table 1) into absolute magnitude, obtaining a value of M$_{B}$$\sim$-20.9 mag. Comparing the B-band luminosity and metallicity of NGC 6845A with respect to the control sample of \citet{kew06}, we found that NGC 6845A display a higher luminosity for its metallicity, as observed for the close pairs studied by \citet{kew06}. This results is consistent with the location of NGC 6845A in the mass-metallicity relation. Therefore, our results are in perfect agreement with the scenario proposed by \citet{kew06}, in the sense that a previous interaction event has produced a gas flow, which transported gas chemically less enriched, lowering the metallicity of this galaxy. 

Although the current data set does not allow us to search for gas motions in this system, we can add all the pieces of evidence in order to explain the observed metallicities. Inspecting Fig. 2 we note that galaxies NGC 6845C and NGC 6845D do not present clear signatures of interactions. In addition, \citet{gordon} did not find HI emission associated with these galaxies. These results suggests that possible glass flows/accretions are not coming from these galaxies. On the other hand, NGC 6845B shows clear signatures of interactions. In addition, \citet{rodrigues} used numerical simulations to study this group, suggesting that NGC 6845A had interacted with NGC 6845B. Therefore, this galaxy seems to be the responsible in producing the dilution of the central metallicity in NGC 6845A. Our current data set allow us to estimate the oxygen abundance of two star-forming regions linked with NGC 6845B (linked by their radial velocities). Both regions (\#1 and \#11) are located in the outskirts of this galaxy and display oxygen abundances of \mbox{12+log(O/H)$\sim$8.4$\pm$0.2}. Therefore, if the galaxy interaction produce a gas flow from the external region of NGC 6845B into the central region of NGC 6845A, it seems reasonable to expect a metal dilution for the central region of NGC 6845A, which before the interaction should have a central abundance of \mbox{12+log(O/H)$\sim$8.8-9.0}, following the mass-metallicity and luminosity-metallicity relations. However, in order to dilute the metallicity of NGC 6845A the amount of less enriched gas should be considerably large. We note that we have not estimated the stellar mass of NGC 6845B, given its perturbed morphology, which could affect the observed colours of this galaxy. In this sense, we are not able to extrapolate the central metallicity of this galaxy, nor estimate their value using the GMOS data, given that their line ratios suggest a possible AGN contribution (as shown in Fig. \ref{bpt}).

Another mechanism to dilute the metallicity of the central region of NGC 6845A could be associated with the accretion and merger of a low-metallicity gas-rich system. Our Gemini imaging data reveals several tidal structures around NGC 6845A. Certainly, some of them should be associated with NGC 6845B. In addition, one star-forming region that is located in the main disk of NGC 6845A (region \#26) display a high H$\alpha$ luminosity \mbox{(L$_{H\alpha}$$\sim$2.4$\times$10$^{40}$ erg s$^{-1}$),} being the most luminous star-forming regions in this system. In this sense, we can not discard that this strong star-forming burst was formed by an accretion event.

As previously found, the metallicity gradient in the tail structures of NGC 6845A is flat. That kind of metallicity distributions has been found in others interacting systems (e.g. NGC 1512, \citealt{bresolin12}, \citealt{lopez15}, NGC 92, \citealt{torres14}). In order to understand the metallicity distribution of the eastern tail structure of NGC 6845A, we can derive a rough estimation of the original metallicity that this galaxy should have had (previous to the interaction) at the optical radius of this galaxy. This exercise can give us an upper limit for the metallicity in the outskirts of NGC 6845A. Considering the stellar mass of NGC 6845A, we can suppose that its central metallicity was about of \mbox{12+log(O/H)$\sim$8.9}, previous to its interaction with NGC 6845B. Now, we can consider the median value of the metallicity gradients derived for \citet{rupkeb} for a sample of galaxies composed by non-interacting systems. In this case, the median slope corresponds to \mbox{-0.57$\pm$0.05 dex/R$_{25}$} (and the zero-points measured for these galaxies are in agreement with the value assumed for NGC 6845A). Using these values, we estimate that the metallicity at R$_{25}$ should be close to \mbox{12+log(O/H)$\sim$8.3}, for the case of NGC 6845A previous to any interaction. Inspecting Fig. \ref{gradiente}, we can note that regions close to R$_{25}$ (regions \#21 and \#22) display oxygen abundances of  \mbox{12+log(O/H)$\sim$8.4$\pm$0.2}. Considering the uncertainties, these measurements are in agreement with the oxygen abundance expected at that radius, for the case of NGC 6845A previous to the interaction (\mbox{12+log(O/H)$\sim$8.3}).

At this point we can suppose that the metallicity gradient of the galaxy extend beyond its optical radius, then, at larger radii we expect lower oxygen abundances. However, we observe a flat metal distribution for the tidal structure, where the abundances are similar to the values measured at the optical radius of NGC 6845A. Therefore, two scenarios can be discussed. On one hand, these results can indicate that the eastern tidal tail was formed from material located in the outskirts of NGC 6845A. This gaseous material was distributed along the tidal tail with a similar chemical content, producing a flat metal distribution, with a similar oxygen abundance than the observed at the optical radius of the galaxy. The dispersion seen in the metal distribution shown in Fig. \ref{gradiente} could be associated with the star formation process along the tail. On the other hand, the tidal tail of NGC 6845A could be originally formed by less enriched material, which increase their metal content due to the star-formation process that is taking place. The main uncertainty here comes from the fact that we do not know the primordial  chemical content of the external region of NGC 6845A. However, we can have an estimation of the enhancement in the oxygen abundance along the tail, which could be produced by star formation. This estimation can be done by using the recipes given by \citeauthor{bresolin12} (\citeyear{bresolin12}, equation 1) and this kind of analysis has been performed recently by \citet{torres14} and \citet{lopez15} for the interacting galaxies NGC 92 and NGC 1512, respectively. In order to do this exercise, we have considered the time on which the star formation took place, which can be associated with the age of the tail (150 Myrs, \citealt{rodrigues}). In addition, we need to consider the HI gas density (which has been taken from \citealt{gordon}) and the star formation rate (SFR) along the tail, which was estimated by using the FUV \textit{GALEX} emission along the tail (measured with the task {\sc polyphot} in {\sc iraf}). The SFR was estimated by using the equations given in \citet{iglesias} and then converted into $\Sigma_{SFR}$. Our estimations give us a \mbox{$\Sigma_{SFR_{FUV}}$=7$\times$10$^{-5}$ M$_{\odot}$ yr$^{-1}$ kpc$^{-2}$} and \mbox{$\Sigma_{HI}$=10.4 M$_{\odot}$ pc$^{-2}$}, for the eastern tidal tail. Finally, we need to know the effective yield \textit{y$_{0}$}. This value is uncertain for star-forming regions located in tidal structures. For instance,  we consider a value of \textit{y$_{0}$}=0.049, which was found by \citet{lopez15} for some external debris in NGC 1512. Considering an initial abundance along the tail of \mbox{12+log(O/H)$\sim$8.2}, the current SFR (derived from UV data) increase the abundance just in 0.01 dex. This result suggests that the star formation along the tail has not been efficient enough to produce an enhancement in its metallicity. As an example, if we increase the $\Sigma_{SFR}$ by one order of magnitude, the oxygen abundance increase just by 0.1 dex. This result can be associated with the large amount of gas present in this location.

So far we have assumed that tidal interaction between NGC6845A and NGC6845B can explain our results. However, we cannot discard that the intruder was a low-mass HI rich galaxy. The galaxy encounter has produced a dilution of the central metallicity of NGC 6845A. Under some assumptions, we found that the star formation in the eastern tidal structure of NGC 6845A has not been highly efficient, which suggest that the star forming regions located in this structure were born from gas already pre-enriched. In this sense, a continuous star formation process since 150 Myrs ago was not enough to increase the oxygen abundances along the tail.

Despite which is the main mechanism responsible in producing the flattening of the metal distribution of NGC 6845A, we note that the use of interacting galaxies in the local Universe can be extremely useful to understand the metallicity gradients observed in the distant Universe, where the interactions were a common phenomenon, but the sources could not well resolved. For instance, \citet{cresci10} studied the chemical distribution in a sample of galaxies at z$\sim$3. These authors suggested that the galaxies in the distant Universe have accreted cold primordial gas (metal-poor gas) and this fact could generate a flat or inverted metallicity gradient. \citet{cresci10} conclude that the observed chemical distribution is due to the metal-poor gas accretion. These authors suggested that the metallicity in the central region of their target has been diluted by accretion of metal-poor gas. In other hand, \citet{sanders15} studied the mass-metallicity relation in the distant Universe (z$\sim$2.3) in a sample of 87 star forming galaxies. They estimated the oxygen abundance using the N2 and O3N2 index. They found that the high-redshift galaxies do not follow the local metallicity relation and they suggest that this results are due to gas inflow rate are higher than SFR and outflow rates.

\subsection{The evolutionary stage of NGC 6845}
\citeauthor{rupkea} (\citeyear{rupkea}a,b) found, through N-body simulations of merging galaxies, that the central abundances in galaxies fall due to inflow of metal-poor gas from the outskirts to the center of the colliding galaxies. Later, \citet{perez} performed simulations considering star formation, supernovae feedback and chemical evolution. These authors found that the central abundances fall slowly after the first encounter and more quickly after the second encounter. Interestingly, we observe a flattening in the metallicity distribution of tidal features belonging to NGC 6845A, which is consistent with the results found by previous authors. However, our data do not allow us to speculate if we are observing the first encounter of NGC 6845A with a companion. On the other hand, \citet{rodrigues} performed numerical simulation to better understand the dynamical evolution of the galaxies that participate in the interaction in this group. These authors found that NGC 6845A probably interacted with NGC 6845B and this event had an age of $\sim$150 Myrs. In addition, using the star-forming regions located in the tail of NGC 6845A, and under some assumptions regarding the kinematic of the tidal structure, these authors derive an age of the collision of $\sim$150-300 Myrs. These estimates suggest a recent encounter. In this sense, the flattening in the metallicity gradient observed on NGC 6845A should complement the previous results obtained by \citet{rodrigues}, suggesting that we are observing the first encounter between NGC 6845A and NGC 6845B (following the results found by \citealt{perez}).

By analyzing the spectra of the central region of each galaxy, as well as using diagnostic diagrams, we found that NGC 6845A, B and D can not be classified as pure star-forming galaxies and there is a potential contribution of an AGN  as a ionization mechanism. Interestingly, several authors have found a connection between AGN activity and interacting galaxies (\citealt{scott} and \citealt{scoville}). In view of our results, we can speculate that the galaxy interactions in the group NGC 6845 can be activating the central black holes of these galaxies, however, integral field spectroscopic observations are necessary to study in details these galaxies.

In addition, galaxy interactions are an efficient mechanism for gas compression and the formation of new stellar complexes (e.g. \citealt{toomre}, \citealt{smith}). In this sense \citet{may} studied star clusters in interacting galaxies. They used 21 cm H{\sc i} maps and the \textit{Hubble Space Telescope} Wide Field Planetary Camera 2. These authors studied the relation between the formation of star clusters and the H{\sc i} column density. The study performed by \citet{may} reveals that the star clusters are formed in regions with \mbox{log(N$_{HI})$$\geq$ 20.6 cm $^{-2}$}. According to the above the H{\sc i} emission in NGC 6845 is \mbox{log(N$_{HI})$$\sim$20.8 cm$^{-2}$} (see \citealt{gordon}), this value is higher than the threshold proposed by \citet{may} for the formation of star clusters in debris of tidal interactions. 

\citet{smith}, \citet{bastian09} and \citet{torres14} have studied the ages of star-forming regions in interacting galaxies (NGC 2856/4, NGC 4038/9 and NGC 92, respectively). These authors found that ages are in the range from $\sim$1 to $\sim$ 30 Myrs. The photometric ages estimated for most star-forming regions in NGC 6845 are $<$ 10 Myrs. These young ages suggest that the star-forming regions studied are formed ``in situ''(\citealt{chien}, \citealt{demelloa}). 

\section{Summary and Conclusions}
We have investigated the star formation and the oxygen abundances in the interacting group NGC 6845 using Gemini GMOS data. This group displays two large tidal tails extending from NGC 6845A to NGC 6845B and NGC 6845A to NGC 6845C. An excitation diagnostic diagram and stellar population synthesis models were used to classify and compute ages for these star-forming regions, which were found to have ages $<$ 10 Myr. We found that the oxygen abundance slightly decreases as a function of distance from the nuclear region of NGC 6845A. 

We suggest that an interaction event has diluted the central metallicity of NGC 6845A. The most probably member involved in the interaction corresponds to NGC 6845B. However, we can not discard the accretion of a low-mass rich HI galaxy. The oxygen abundance measured along the tail can not be explained by a continuous star formation process, given the low $\Sigma_{SFR}$. The observed level of star formation could produce a small enhancement in the abundance, which can be associated with the large amount of HI that lies at the location of the eastern tidal tail. In this sense, the oxygen abundances measured along the tail structure  suggest that this region was contaminated with pre-enriched gas (during the interaction with NGC 6845B). From this gas the star-forming regions that we observe on this tail were born.

The complex nature of this interacting system shows the need for a large multi wavelength survey of different types of interacting systems to disentangle the importance of tidal effects in shaping up galaxies properties.

\section{ACKNOWLEDGEMENTS}
We thank the anonymous referee for the useful comments that greatly improved this paper. DO-R acknowledges the financial support of the Direcci\'{o}n de Investigaci\'{o}n of the Universidad de La Serena, through a ``Concurso de Apoyo a Tesis 2013'', under contract PT13145. ST-F acknowledges the financial support of the Chilean agency FONDECYT through a project ``Iniciaci\'{o}n en la Investigaci\'{o}n'', under contract 11121505 and also the financial support of the project CONICYT PAI/ACADEMIA 7912010004. DdM thanks FONDECYT and the hospitality of the Universidad de la Serena. CMdO aknowledges support from FAPESP and CNPq. This research was based on observations obtained at the Gemini Observatory South , which is operated by the Association of Universities for Research in Astronomy, Inc., under a cooperative agreement with the NSF on behalf of the Gemini partnership: the National Science Foundation (US), the Science and Technology Facilities Council (UK), the National Research Council (Canada), CONICYT (Chile), the Australian Research Council (Australia), Minist\'{e}rio da Ci\^{e}ncia e Tecnologia (Brazil) and Ministerio de Ciencia, Tecnolog\'{i}a e Innovaci\'{o}n Productiva (Argentina). This research has made use of the NASA/IPAC Extragalactic Database (NED) which is operated by the Jet Propulsion Laboratory, California Institute of Technology, under contract with the National Aeronautics and Space Administration. \textit{GALEX} is a NASA Small Explorer, launched in 2003 April. We gratefully acknowledge NASA's support for construction, operation and science analysis for the GALEX mission, developed in cooperation with the Centre National d’ Etudes Spatiales of France and the Korean Ministry of Science and Technology. We also acknowledge the use of the HyperLeda data base (\url{http://leda.univ-lyon1.fr})	

\bibliographystyle{mn2e}
\bibliography{biblo.bib}
			
\clearpage

\end{document}